\documentclass[twocolumn,preprintnumbers,amsmath,amssymb]{revtex4}
\usepackage{graphicx}% Include figure files
\usepackage{dcolumn}% Align table columns on decimal point
\usepackage[utf8x]{inputenc}
\usepackage{bm}% bold math

\begin{document}

\title{Spin Averaged Mass Spectrum of Heavy Quarkonium via Asymptotic Iteration Method}% Force line breaks with \\

\author{Halil Mutuk}
\email{halilmutuk@gmail.com}
 \affiliation{Physics Department, Ondokuz Mayis University, 55139, Samsun, Turkey}%Lines break automatically or can be forced with \\

\date{\today}% It is always \today, today,

\begin{abstract}
In this paper we solved Schrödinger equation with Song-Lin potential by using Asymptotic Iteration Method (AIM). We obtained spin-averaged energy levels of charmonium and bottomonium via AIM. Obtained results agree well with available experimental data and other theoretical studies. 
\end{abstract}

\pacs{Valid PACS appear here}% PACS, the Physics and Astronomy
\maketitle

\section{\label{sec:level1}Introduction}

Quarkonium is the bound state of a quark and antiquark by the strong interaction. Excluding top quark $t$, the rest of the quarks $b,c,u,d,s$ can make quarkonium states. Large masses of $b$ and $c$ quarks comparing to others deserve additional interest since experimentally $b\bar{b}$ and $c\bar{c}$ systems give fruitful data in the experiments.  

The universe that we observe is mainly made of hadrons (mesons + baryons). Hadrons are strongly interacting particles that are made of quarks. The interaction of quarks is described by Quantum Chromodynamics (QCD), which is part of the Standart Model of particle physics. QCD is thought to be the {\it true} theory of strong interactions. QCD is an SU(3) gauge theory describing the interactions of six quarks which transform under the fundamental representation of SU(3) group via the exchange of gluons that transform under the adjoint representation. Although it has been more than 50 years that QCD has been proposed, it has evaded a solution. Contrary to electroweak theory, where it is possible to obtain precise results using perturbation theory, the order of precision obtained in QCD has been lower by orders of magnitude. The main reason for this is that the coupling constant, which should be the perturbation parameter, of QCD is of the order one in low energies, hence the truncation of the perturbative expansion can not be carried out. However, it is an important subject to study the spectrum of particles predicted by QCD.

In potential quark models one assumes a potential interaction, intrinsically it is a non-relativistic approach. Therefore, the systems that are best suited for study in quark models are the heavy quark systems that contain $c$ or $b$ quarks. Quark potential models gave considerable results in accounting for the observed spectroscopy of heavy quarkonium \cite{1}. There are many phenomenological potentials to study heavy quarkonium physics. These potentials can be classified as phenomenological (non-QCD like) and QCD motivated (or QCD inpired) potentials. For example,  Logarithmic \cite{2}, Richardson \cite{3},  Buchmüller-Tye \cite{4} potentials are phenomenological spin-independent potentials and the potential in \cite{5} is spin-dependent and velocity-dependent. The nature of quarkonium potentials are somehow not clear. $q\bar{q}$ potentials cannot be derived from first principles of QCD. Therefore there is no dominant potential which interpret quarkonium spectrum exactly. There are nonrelativistic quark models \cite{6,7} and relativistic models \cite{8,9,10} to study spectrum. A general review can be found in \cite{11} and \cite{12} and references therein. In \cite{13}, C. Hong et al. studied properties of heavy quarkonium systems using Dick potential which explicitly included the color factor in the potential. 

The Song-Lin potential was presented in \cite{14} as a new phenomenological potential
for heavy quarkonium:
\begin{equation}
V(r)=-br^{(-1/2)}+ar^{(1/2)}
\end{equation} 
where a and b are adjustable parameters. The potential has a Lorentz vector term motivated by experimental leptonic widths for vector mesons and perturbative QCD at short distances and a Lorentz scalar term responsible for quark confinement at large distances. Song-Lin potential was suggested to describe heavy quarkonium spectra in the late '90s. Being not a directly QCD inspired potential, it explained well energy levels, leptonic decay widths, radiative transition rates and hadronic decay rates. They obtained spin averaged mass spectrum by solving two body Shrödinger equation numerically \cite{14}.

This work is devoted to obtain eigenvalues of the radial Schrödinger equation with the Song-Lin potential

\begin{equation}
\frac{d^2 R_{nl}(r)}{dr^2}+\frac{2m}{\hbar^2}\left( E-V(r)-\frac{l(l+1)}{2mr^2} \right) R_{nl}(r)=0, \label{eqn11}
\end{equation}
 where $m$ is the mass of the particle. Inserting Song-Lin  potential into Eqn. (\ref{eqn11}) radial Schrödinger equation takes the following form:
 
 \begin{equation}
\frac{d^2 R_{nl}(r)}{dr^2}+\frac{2m}{\hbar^2}\left( E+br^{(-1/2)}-ar^{(1/2)}-\frac{l(l+1)}{2mr^2} \right) R_{nl}(r)=0 \label{eqn2}.
\end{equation}

In order to obtain spin averaged mass spectrum for heavy quarkonium systems, the above equation is solved via Asymptotic Iteration Method (AIM). This method was derived in \cite{15} and generalized for exactly solvable eigenvalue problems in \cite{16}.

\section{\label{sec:level2}Formalism of the Problem}
The asymptotic iteration method was introduced to  solve second-order homogeneous linear differential equations of the form

\begin{equation}
y^{\prime \prime}(x)=\lambda_0(x) y^\prime (x) + s_0(x) y_0(x) \label{eqn3}
\end{equation}

where $\lambda_0(x)\neq 0$ and the variables $\lambda_0(x)$ and $s_0(x)$ have sufficiently many derivatives \cite{15}. This differential equation has a solution as follows:

\begin{eqnarray}
y(x)&=&\texttt{exp} \left( -\int^x \alpha dt \right)\nonumber \\ & \times & \left[ C_2+C_1 
 \int^x \texttt{exp} (\int^t \left[(\lambda_0(\tau)+2\alpha(\tau))d\tau \right]   dt  \right] \label{eqn4}
\end{eqnarray}
 where 
\begin{equation}
\frac{s_k(x)}{\lambda_k(x)}=\frac{s_{k-1}(x)}{\lambda_{k-1}(x)} \equiv \alpha(x). \label{eqn5}
\end{equation}
for sufficiently large $k$.  In Eqn. (\ref{eqn5}) $s_k(x)$ and $\lambda_k(x)$ are defined as follows:

\begin{eqnarray}
\lambda_k(x)=\lambda^\prime_{k-1}(x)+s_{k-1}(x)+\lambda_0(x) \lambda_{k-1}(x), \nonumber \\
s_n(x)=s^\prime_{k-1}(x)+s_0(x)\lambda_{k-1}(x),~ k=1,2,3, \cdots \label{eqn6}
\end{eqnarray}

The convergence (quantization) condition can be defined as 
\begin{equation}
\delta_k(x)=\lambda_k(x)s_{k-1}(x)-\lambda_{k-1}(x)s_k(x)=0, ~ k=1,2,3, \cdots  \label{eqn7}
\end{equation}
For a given radial potential, it is possible to convert radial Schrödinger equation into Eqn. (\ref{eqn3}). Once this form has been obtained, it is easy to see $s_0(x)$ and $\lambda_0(x)$ and calculate  $s_k(x)$ and $\lambda_k(x)$ by using Eqn. (\ref{eqn6}). Eigenvalues are obtained from the quantization (convergence) condition, $\delta_k(x)=0$. 

\subsection{AIM for Song-Lin Potential}

Defining $\frac{2mE_n}{\hbar^2}=\epsilon$, $\frac{2mb}{\hbar^2}=\tilde{b}$ and $\frac{2ma}{\hbar^2}=\tilde{a}$, Eqn. (\ref{eqn2}) can be written as follows
\begin{equation}
\frac{d^2 R_{nl}(r)}{dr^2} + \left\lbrace \epsilon +\tilde{b}r^{-1/2}-\tilde{a}r^{1/2} +\frac{l(l+1)}{r^2} \right\rbrace R_{nl}(r)=0 \label{eqn8}.
\end{equation}
This is a second order linear differential equation and in order to solve this equation via AIM, we should transform it into Eqn. (\ref{eqn3}). For this we use a wave function of the form

\begin{equation}
R_{nl}(r)=r^{l+1} \texttt{exp}(-\alpha r^2)f_{nl}(r) \label{eqn9}.
\end{equation}

Inserting this wave function in Eqn. (\ref{eqn8}) we have second order linear homogeneous differential equation like

\begin{eqnarray}
\frac{d^2 f_{nl}(r)}{dr^2}&=&\left( -\frac{2 (l+1)}{r} + 4 \alpha  r \right) \frac{df_{nl}(r)}{dr} \nonumber \\  &+& [-\epsilon -2 \tilde{b}r^{-1/2}+2\tilde{a}r^{1/2} \nonumber \\ 
&+&6\alpha +4 l \alpha -4r^2 \alpha]  f_{nl}(r) \label{eqn10}.
\end{eqnarray}
By comparing Eqn. (\ref{eqn10}) with Eqn. (\ref{eqn3}), we see that  AIM can be applied and $\lambda_0(x)$ and $s_0(x)$ can be written as follows:

\begin{equation}
\lambda_0(x)=\left( -\frac{2 (l+1)}{r} + 4 \alpha  r \right)
\end{equation}
and 
\begin{equation}
s_0(x)= [-\epsilon -2 \tilde{b}r^{-1/2}+2\tilde{a}r^{1/2} +6\alpha +4 l \alpha -4r^2 \alpha].
\end{equation}

Armed with these, one can obtain spin averaged heavy quarkonium spectra. The results are shown in Table \ref{tab:table1} and Table \ref{tab:table2} for charmonium and bottomonium, respectively. We used constituent quark masses as $m_c=1.80 ~ \text{GeV} $, $m_b= 5.20 ~ \text{GeV}$, $a=1.15 ~ \text{GeV}~ \text{fm}^{-1/2} $ and  $b=0.41 ~ \text{GeV}~ \text{fm}^{1/2} $ \cite{14}. 

\begin{table}[h]
\caption{\label{tab:table1}Charmonium mass spectrum (in MeV). }
\begin{ruledtabular}
\begin{tabular}{cccccccccc}
 State &Exp. \cite{17}& This work & \cite{14} & \cite{3}&
 \cite{4}& \\
\hline
1S& 3067  & 3104 & 3097 & 3095 & 3100 &\\
1P& 3525 & 3572 & 3524 & 3514 & 3520 &\\
2S& 3649 & 3703 & 3672 & 3684 & 3700 & \\
1D& 3769 & 3806 & 3791 & 3799 & 3810 & \\
2P&      & 3986 & 3907 & 3950 & 3970 & \\ 
3S& 4040 & 4090 & 4017 & 4096 & 4120 & \\
2D& 4159 & 4185 & 4090 & 4172 & 4190 & \\
3P&      & 4280 & 4186 & 4308 &      & \\
4S& 4415 & 4375 & 4275 & 4440 & 4480 & \\
3D&      & 4474 & 4328 &      &      & \\
4P&      & 4580 & 4409 &      &      & \\
5S&      & 4692 & 4487 &      &      & \\
\end{tabular}
\end{ruledtabular}

\end{table}

\begin{table}[h]
\caption{\label{tab:table2}Bottomonium mass spectrum (in MeV). }
\begin{ruledtabular}
\begin{tabular}{cccccccccc}
 State &Exp. \cite{17}& This work & \cite{14} & \cite{3}&
 \cite{4}& \\
\hline
1S& 9444  & 9473  & 9460  & 9452  & 9460  &\\
1P& 9900  & 9912  & 9902  & 9888  & 9890  &\\
2S& 10023 & 10024 & 10034 & 10007 & 10020 &\\
1D& 10161 & 10156 & 10162 & 10137 & 10140 &\\
2P& 10260 & 10275 & 10261 & 10241 & 10250 &\\ 
3S& 10355 & 10327 & 10356 & 10338 & 10350 &\\
2D&       & 10434 & 10432 & 10421 & 10430 &\\
3P&       & 10580 & 10512 & 10512 & 10530 &\\
4S& 10579 & 10593 & 10589 & 10598 & 10620 &\\
3D&       & 10625 & 10643 &       & 10680 &\\
4P&       & 10703 & 10711 &       &       &\\
5S& 10865 & 10788 & 10777 &       & 10860 &\\
\end{tabular}
\end{ruledtabular}

\end{table}

\section{Discussion and Conclusion}
In this paper, we studied Song-Lin potentital which is a phenomenological potential by AIM. AIM is an iterative method and converges fairly good. Given a potential, one can find energy eigenvalues for any $n$ and $l$. This method can be applied to other potentials in the literature. To the best of our knowledge, we first solved Schrödinger equation with Song-Lin potential by using this method. We obtained spin averaged mass spectra of charmonium $c\bar{c}$ and bottomonium $b\bar{b}$. Obtained results agree well with experimental data and other studies. Song-Lin potential explains well spin-averaged mass spectra of heavy quarkonium. For a complete picture, one must add relativistic corrections in the calculation of energy levels. This procedure results splittings in the energy levels. Most of the quark model potentials, i.e., phenomenological potentials are similar. They have a Coulomb like term and a linear term. Relativistic corrections, spin-spin or spin-orbit interactions can be added to the potential in a particular study. These effects in general are small compared to the given potential. So without these considerations one can find reliable results compatible with the experimental data.

\end{document}